\documentclass{elsart}
\usepackage{graphicx}
\usepackage{epsfig}
\usepackage{amssymb}
\begin{document}

\begin{frontmatter}

\title{
Four Quark $cn$-$\bar n\bar c$ States in $\tilde U(12)$-Scheme and\\ 
$X(3872)/Y(3940)$
}

\author{Muneyuki Ishida$^a$, Shin Ishida$^b$ and Tomohito Maeda$^c$
}
\address{$^a$Department of Physics, School of Science and Engineering, Meisei University, 
Hino, Tokyo 191-8506, Japan}
\address{$^b$Research Institute of Science and Technology, 
College of Science and Technology,\\
Nihon University, Tokyo 101-8308, Japan}
\address{$^{c}$ Department of Engineering Science, Junior College Funabashi Campus,\\
Nihon University, Funabashi 274-8501, Japan}

\begin{abstract}

The properties of four quark $cn$-$\bar n\bar c$ states
are investigated as $cn$ di-quark and $\bar n \bar c$ di-antiquark system
in $\tilde U(12)$-classification scheme of hadrons, recently proposed by us.
We consider the negative-parity di-quark and di-antiquark in ground states, 
which form with the ordinary positive-parity ones the respective 
linear representations of chiral symmetry.
The masses of ground-states are predicted by using Joint Spring Quark Model, and 
the observed properties of $X(3872)$ and $Y(3940)$ are consistent, respectively, with those 
of the $J^{PC}=1^{++}$ and $2^{++}$ states from the negative-parity di-quark and di-antiquark.
Their narrow-widths are explained from an orthogonality of spinor wave functions.
The properties of ground-state $cs$-$\bar s \bar c$ system are also predicted in this scheme.  

\end{abstract}

\end{frontmatter}

\section{Introduction}
\label{sec1}

The discovery of $X(3872)$\cite{X} and $Y(3940)$\cite{Y} by Belle presented new problems on 
hadron spectroscopy. Because of their decay-properties,
they are considered to be non-$c\bar c$ states, 
and there are some possible explanations for the origin of these states:
$D^0\bar D^{*0}$ molecule state\cite{Swanson} and
four-quark $cn$-$\bar n\bar c$ state\cite{polosa}.
In this letter, we consider the four-quark explanation of $X(3872)/Y(3940)$, where 
the relevant particles are considered as the $cn$ di-quark and $\bar n \bar c$ di-antiquark system.
This explanation is criticized from the observed decay widths.
The $cn$-$\bar n \bar c$ system decays to $D\bar D^{(*)}$ and $J/\psi\rho$(or $J/\psi\omega)$.
They are so-called OZI super-allowed processes, 
and proceed without no $q\bar q$-pair production or annihilation. 
Their widths are expected to be so large as a few hundred MeV or more.
On the other hand the observed widths of $X(3872)$ and $Y(3940)$ are very small, 
$\Gamma_X < 2.3$MeV\cite{X}\footnote{
The other (and main) reason of this tiny $\Gamma_X$ is explained from its small phase space
of the decays to $D^0\bar D^{*0}$ and $J/\psi\ \rho(\rightarrow\pi\pi)$.
} 
and $\Gamma_Y=87\pm 22\pm 26$MeV\cite{Y}, respectively.

A few years ago\cite{[1]}, we have proposed a covariant level-classification scheme of hadrons
based on $\tilde U(12)$ spin-flavor group. It is a relativistic generalization of the 
$SU(6)_{SF}$-scheme in non-relativistic quark model(NRQM), and the squared-mass spectra of hadrons including 
light constituent quarks are classified as the representations of $\tilde U(12)$. 
(Concerning the $\tilde U(12)$-scheme and its group theoretical arguments,
 see our ref.\cite{[2]}.)
The essential difference from $SU(6)_{SF}$ is a spinor WF.
Each spinor index corresponding to light constituent quark in composite hadron WF is expanded by 
free Dirac spinors $u_{r,s}(v_\mu )$ called urciton spinors. At the rest frame 
$u_{+,s}({\bf 0} )$ has the upper two-component which corresponds to the Pauli spinor 
appearing in NRQM, while the $u_{-,s}({\bf 0} )$ has the lower two-component
representing the relativistic effect.
The $r$ index of $u_{r,s}$ represents the eigen-value of $\rho_3$, and the corresponding freedom
is called $\rho$-spin, while the ordinary Pauli-spin freedom is called $\sigma$-spin.
In the $\tilde U(12)$-scheme, 
the $u_{-,s}$ is supposed to appear as new degrees of freedom for light constituents,
being independent of $u_{+,s}$,
and we predict the existence of a number of extra states out of the $SU(6)_{SF}$-framework
for meson and baryon systems. These extra states are called chiral states, since
$u_{-,s}$ is obtained from $u_{+,s}$ through chiral $\gamma_5$ transformation.
The $\tilde U(12)$ representation is considered to be actually realized in light-quark baryon spectra.
The ground-state $qqq$ baryons, $N$-octet and $\Delta$-decouplet,
have positive-parity. Experimentally, the next low-lying states, 
$N(1440)$, $\Lambda (1600)$, $\Delta (1600)$, $\cdots$, have also positive-parity,
contrarily to the expectation of NRQM. In $\tilde U(12)$-scheme this parity-doubling problem
is resolved in its representation. 
The $qqq$-baryons and antibaryons in ground states form {\bf 364}, which includes
two {\bf 56} baryons with positive-parity in terms of $SU(6)_{SF}$.
The extra {\bf 56} has $\rho_3$ WF, $(+,-,-)$.
The masses and strong decay widths of positive-parity baryons,
$N(1440)$, $\Lambda (1600)$ and $\Sigma (1660)$, 
are consistent with this extra {\bf 56}\cite{baryon}, while 
the ordinary radially excited states have large widths of several hundred MeV
and are expected not to be observed as resonant particles. 

The heavy-light $Q\bar q$ mesons in ground states are classified as ${\bf 12}^*$, 
which includes the chiral $J^P=0^+/1^+$ states with positive-parity 
as well as the ordinary $0^-/1^-$ states.
They form linear representations of chiral symmetry\cite{[2],Bardeen}.
The chiral states have $(\rho_3^Q,\rho_3^{\bar q})=(+,-)$, and they are expected to 
appear in smaller mass region as different states from the $P$-wave states.
The masses and decay properties of $D_s(2317)/D_s(2460)$ are naturally explained as
the chiral states in $c\bar s$ system, where $\bar s$ has negative $\rho_3$ quantum number.
Applying the $\tilde U(12)$-scheme,
the heavy-light $Qq$ di-quark (light-heavy $\bar q\bar Q$ di-antiquark) forms ${\bf 12}({\bf 12}^*)$
in ground states, which includes chiral $J^P=0^-/1^-$ states with negative parity as well as the 
ordinary $0^+/1^+$ states. The chiral $Qq(\bar q\bar Q)$ states have negative 
$\rho_3^q(\rho_3^{\bar q})$, and they are out of the non-relativistic framework.

We have pointed out in our previous work\cite{baryon,SIshida} 
a kind of phenomenological conservation rule applicable to chiral states, called $\rho_3$-line rule.
In transitions of chiral states, the $\rho_3$ quantum number along each spectator-quark line is 
approximately conserved. The violation of $\rho_3$ is suppressed by the factor ${\bf P}/2M$,
where ${\bf P}(M)$ is the final hadron momentum(mass), 
(not by ${\bf P}/2m_q$, where $m_q$ is the constituent quark mass).

There is an interesting possibility that the observed $X(3872)$ and $Y(3940)$ are chiral states,
which are made from negative-parity $cn$ di-quark and $\bar n\bar c$ di-antiquark, 
and that their narrow-width properties are explained by 
the $\rho_3$-line rule, 
since the relevant decays are quark-rearrangement processes and proceed only through 
the spectator-quark interaction.
The purpose of this letter is to investigate the masses and properties of strong decays of
ground $cn$-$\bar n\bar c$ states in $\tilde U(12)$-scheme, and to consider
whether the $X(3872)$ and $Y(3940)$ are explained as chiral $cn$-$\bar n\bar c$ states or not.

\section{Di-quark and Di-antiquark Spinor Wave Function}

As was explained in the previous section, in $\tilde U(12)$-scheme
the negative-parity $cn$ di-quark($\bar n\bar c$ di-antiquark), called chiral state, 
appears in ground-state. 
They are denoted as $\Gamma_{cn}^\chi (\Gamma_{\bar n \bar c}^\chi )$, which has
negative $\rho_3^n(\rho_3^{\bar n})$, while the ordinary
positive-parity di-quark(di-antiquark), called Pauli state, is denoted as 
$\Gamma_{cn}^P(\Gamma_{\bar n \bar c}^P)$,
which is composed of constituents both with positive $\rho_3$.
Their explicit forms are given in Table \ref{tab1}. 
\begin{table}
\begin{center}
\begin{tabular}{cc|c||cc|c}
$(\rho_3^c,\rho_3^n)$ & $J^P$ & $\Gamma_{cn}^{P}$(Pauli state) 
 & $(\rho_3^c,\rho_3^n)$ & $J^P$ & $\Gamma_{cn}^{\chi}$(chiral state) \\
\hline
$(+,+)$ & $0^+$ & $(1-iv\cdot\gamma) \frac{-\gamma_5}{2\sqrt 2} C^\dagger$ 
 & $(+,-)$ & $0^-$ & $(1-iv\cdot\gamma) \frac{1}{2\sqrt 2}C^\dagger$ \\
$(+,+)$ & $1^+$ & $(1-iv\cdot\gamma) \frac{i\epsilon\cdot\gamma}{2\sqrt 2}C^\dagger$ 
 & $(+,-)$ & $1^-$ & $(1-iv\cdot\gamma) \frac{i\gamma_5\epsilon\cdot\gamma}{2\sqrt 2}C^\dagger$ \\
\hline\hline
$(\rho_3^{\bar n},\rho_3^{\bar c})$ & $J^P$ & $\Gamma_{\bar n \bar c}^{P}$(Pauli state) 
 & $(\rho_3^{\bar n},\rho_3^{\bar c})$ & $J^P$ & $\Gamma_{\bar n \bar c}^{\chi}$(chiral state) \\
\hline
$(+,+)$ & $0^+$ & $C^\dagger \frac{-\gamma_5}{2\sqrt 2} (1+iv\cdot\gamma) $ 
 & $(-,+)$ & $0^-$ & $C^\dagger \frac{-1}{2\sqrt 2} (1+iv\cdot\gamma)$ \\
$(+,+)$ & $1^+$ & $C^\dagger \frac{i\epsilon\cdot\gamma}{2\sqrt 2}(1+iv\cdot\gamma) $ 
 & $(-,+)$ & $1^-$ & $C^\dagger \frac{i\gamma_5\epsilon\cdot\gamma}{2\sqrt 2}(1+iv\cdot\gamma) $ \\
\hline
\end{tabular}
\end{center}
\caption{Di-quark and Di-antiquark spinor WF $\Gamma_{cn}$ and $\Gamma_{\bar n \bar c}$ 
in $\tilde U(12)$-scheme. The $\rho_3$ WF are also shown.
The charge conjugation matrix $C^\dagger =\gamma_2\gamma_4$.
}
\label{tab1}
\end{table}

As is seen from Table \ref{tab1}, aside from $C^\dagger$, the $\Gamma_{cn}$ 
is the same as the WF of $c\bar n$-meson $H_{c\bar n}$ in heavy quark effective theory.
That is, $\Gamma_{cn} = H_{c\bar n} C^\dagger$. 
The $0^+/1^+$ di-quark WF correspond to $0^-/1^-$ meson WF, and the $0^-/1^-$ chiral di-quark WF
correspond to the $0^+/1^+$ chiral meson WF. 
The chiral $0^+/1^+$ mesons become heavier\cite{[2],Bardeen} than the $0^-/1^-$ states 
through spontaneous chiral symmetry breaking($\chi$SB),
which is described by the Yukawa coupling to the scalar $\sigma$-meson nonet 
$s=s^i\lambda^i/\sqrt 2$ in the framework of $SU(3)$ linear $\sigma$ model\cite{LsM}.
In $Qq$ di-quark systems, because of the $C^\dagger$ factor, the chiral $0^-/1^-$ states 
are expected to have smaller masses than the $0^+/1^+$ Pauli states. 
The mass shift $\delta M^{\chi ,P}$ in $\chi$SB is given by
\begin{eqnarray}
\delta M^{\chi ,P} &=& g_\sigma s_0\ 
           \langle \Gamma_{cn}^{\chi ,P} \overline{\Gamma_{cn}^{\chi ,P}} \rangle
     = g_\sigma s_0\ 
           \langle H_{c\bar n}^{\chi ,P} C^\dagger C^\dagger 
                   \overline{H_{c\bar n}^{\chi ,P}} \rangle
     = - g_\sigma s_0\ \langle H_{c\bar n}^{\chi ,P} \overline{H_{c\bar n}^{\chi ,P}} \rangle ,
\ \ \ \ \ \ \ \ 
\label{eq1}
\end{eqnarray}
where $\overline{\Gamma_{cn}^{\chi ,P}}\equiv \gamma_4 (\Gamma_{cn}^{\chi ,P})^\dagger \gamma_4$, and 
the change of sign comes from $(C^\dagger)^2=-1$.
The $s_0$ is vacuum expectation value of $s$ and $g_\sigma$ is the Yukawa coupling constant.
The chiral mass splitting $\Delta M_\chi (\equiv \delta M^P-\delta M^\chi ) =2g_\sigma s_0$ 
is predicted with 242MeV in our previous work\cite{[2]},
which is estimated by using the masses of $D_s(2317)$ and $D_s(2460)$ as inputs.   

Concering $cn$-$\bar n\bar c$ system, 
we expect three combinations of $\Gamma_{cn}$-$\Gamma_{\bar n\bar c}$:
The $\Gamma_{cn}^P$-$\Gamma_{\bar n\bar c}^P$ denoted as Pauli-Pauli states,
the $\Gamma_{cn}^P$-$\Gamma_{\bar n\bar c}^\chi$
(or $\Gamma_{cn}^\chi$-$\Gamma_{\bar n\bar c}^P$) denoted as Pauli-chiral(or chiral-Pauli) states, 
and the $\Gamma_{cn}^\chi$-$\Gamma_{\bar n\bar c}^\chi$ denoted as chiral-chiral states.
Among them the chiral-chiral states have the smallest masses, the chiral-Pauli states are heavier
by $\Delta M_\chi$, and the Pauli-Pauli states are further heavier by $\Delta M_\chi$.
For the transitions between chiral and Pauli states, the amplitudes
of pion emission are expected to be very large, since   
the pseudoscalar Nambu-Goldstone bosons have the spinor WF $i\gamma_5/2$, which changes
$\rho_3$ quantum number of the interacting quark.
The Pauli-Pauli states decay to Pauli-chiral states through pion emission, and 
the Pauli-chiral states do to chiral-chiral states.
These decays are $S$-wave, and the widths are expected to be so large as several hundred MeV or a GeV.
Accordingly, the Pauli-Pauli states and Pauli-chiral states are not observed
as resonances, but as non-resonant backgrounds.

This is quite similar mechanism to the situation in ${\bf 70}_G$ baryons in $\tilde U(12)$-scheme.
They are not observed as resonances because of their large widths of $\pi$ (or $K$) decays 
to ${\bf 56}_E$ baryons, except for $\Lambda (1405)$\cite{baryon}.
We focus our interests on chiral-chiral states in the following.
Their masses are predicted by considering spin-spin interaction in the next section.

\section{Oscillator WF and Ground-State Mass Spectra}

\begin{figure}[t]
  \epsfxsize=10 cm
  \epsfysize=7 cm
 \centerline{\epsffile{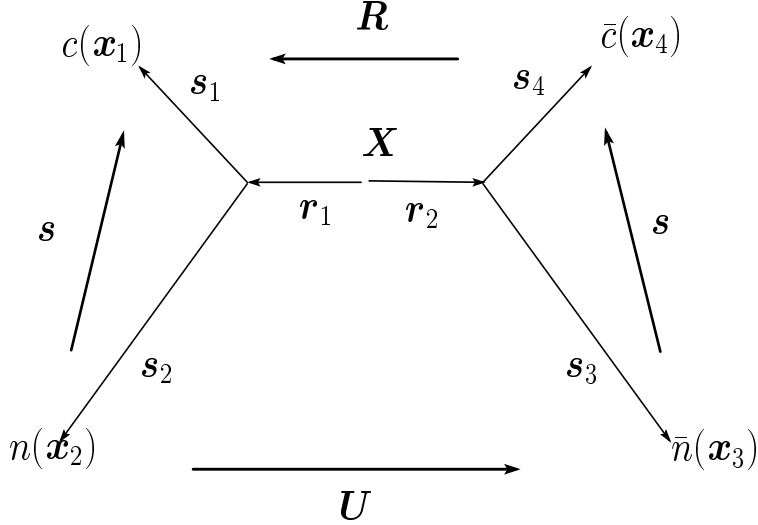}}
 \caption{Space coordinates in $cn$-$\bar n\bar c$ system.
          Eigen-modes \mbox{\boldmath $s$}, \mbox{\boldmath $R$}, \mbox{\boldmath $U$} 
          are also depicted.}
  \label{fig1}
\end{figure}

({\it Oscillator WF})\ \ \ 
In order to predict the mass spectra of ground states,
it is necessary to estimate quantitatively the spin-spin interaction.  
The space WF of $cn$-$\bar n \bar c$ system are required for this purpose, and
we use the joint spring quark model(JSQM)\cite{JSQM},
where the $cn$ di-quark ($\bar n\bar c$ di-antiquark) is in color ${\bf 3}^*({\bf 3})$,
and they are connected by spring potential as shown in Fig.~\ref{fig1}.
%
%
Correspondingly, the Lagrangian of the four quark system is given by
\begin{eqnarray}
L &=& \frac{m_c}{2}(\dot{\bf x}_1^2+\dot{\bf x}_4^2) + \frac{m_n}{2}(\dot{\bf x}_2^2+\dot{\bf x}_3^2)
    -\frac{k}{2} ( {\bf s}_1^2+{\bf s}_2^2+{\bf s}_3^2+{\bf s}_4^2 )
    -\frac{k}{2} ( {\bf r}_1 - {\bf r}_2 )^2\ ,\ \ \ \ \ 
\label{eq3}
\end{eqnarray}
where ${\bf x}_{1,2}={\bf X}+{\bf r}_1+{\bf s}_{1,2}$ and 
${\bf x}_{3,4}={\bf X}+{\bf r}_2+{\bf s}_{3,4}$. 
The definition of each space coordinate
is given in Fig.~\ref{fig1}. The $m_c(m_n)$ are constituent quark mass of $c(n)$ and 
$k$ is the spring constant.  
The analytic formulas for eigen-modes are somewhat complicated, so we show only numerical results.
\begin{eqnarray} 
{\bf s} &=& ({\bf x}_1-{\bf x}_2-{\bf x}_3+{\bf x}_4)/\sqrt 2,\ \ 
{\bf R}  = 0.785 ({\bf x}_1-{\bf x}_4) + 0.083 ({\bf x}_2-{\bf x}_3),\nonumber \\ 
{\bf U} &=& 0.365 ({\bf x}_1-{\bf x}_4) - 0.862 ({\bf x}_2-{\bf x}_3),\ \ 
\label{eq4}
\end{eqnarray}
where we take $m_c=M_{J/\psi}/2=1.548$GeV and $m_n=m_\rho /2=0.385$GeV.
The $k$ is determined from the inverse Regge slope of $n\bar n$ meson system, 
$\Omega_{n\bar n}=\sqrt{32 m_n k}=1.14$GeV$^2$, as $k=0.107$GeV$^3$\cite{Regge}. 
We may regard {\bf s} as the relative coordinate in di-quark or di-antiquark,
{\bf R}({\bf U}) approximately as relative coordinate between $c$ and $\bar c$($n$ and $\bar n$). 
The internal space WF is given by
\begin{eqnarray}
f_s({\bf s}) f_R({\bf R}) f_U({\bf U}) &=& 
\left( \frac{\beta_s\beta_R\beta_U}{\pi^3} \right)^{\frac{3}{4}} e^{-\frac{\beta_s}{2}{\bf s}^2}
e^{-\frac{\beta_R}{2}{\bf R}^2}e^{-\frac{\beta_U}{2}{\bf U}^2},
\label{eq5}
\end{eqnarray}
where $\beta_s=0.129$GeV$^2$, $\beta_R=0.254$GeV$^2$ and $\beta_U=0.115$GeV$^2$.\\
({\it Spin-spin interaction and Mass spectra})\ \ \ 
By using the WF (\ref{eq5}), we can estimate the magnitudes of spin-spin interactions,
\begin{eqnarray} 
H_{SS} &=& -\frac{\alpha_{eff}}{m_i m_j} \delta^{(3)} ({\bf x}_i-{\bf x}_j) 
\mbox{\boldmath $\sigma^{(i)}$}\cdot \mbox{\boldmath $\sigma^{(j)}$}\ C_{\rm Casimir}\ . 
\label{eq6}
\end{eqnarray}
The effective coupling $\alpha_{eff}$ is determined from the spin-spin splitting of $D$-meson system,
$m_{D^*}-m_D= (\alpha_{eff}/(m_cm_n))    (\beta /\pi)^{3/4} 4\cdot (4/3)=142$MeV as $\alpha_{eff}=1.14$,
where the extension parameter of $D$ system is 
$\beta = \sqrt{\frac{m_cm_nk}{m_c+m_n}}=0.181$GeV$^2$.
Now we can predict the strengths of spin-spin interactions in all possible pairs 
of constituents. The results are given in Table \ref{tab2}.  
\begin{table}
\begin{tabular}{l|cc|l}
   & $\langle \delta^{(3)}({\bf x}_i-{\bf x}_j) \rangle $ & -$C_{\rm Casimir}$ 
         & $\langle -C_{\rm Casimir} \frac{\alpha_{eff}}{m_im_j} \delta^{(3)}({\bf x}_i-{\bf x}_j) 
            \rangle \mbox{\boldmath $\sigma^{(i)}$}\cdot \mbox{\boldmath $\sigma^{(j)}$}$ \\
\hline
$c_1$-$n_2$, $\bar n_3$-$\bar c_4$ & 0.0085GeV$^3$
 & $\frac{2}{3}$ & 10.9MeV$( \mbox{\boldmath $\sigma^{(1)}$}\cdot \mbox{\boldmath $\sigma^{(2)}$}
                   + \mbox{\boldmath $\sigma^{(3)}$}\cdot \mbox{\boldmath $\sigma^{(4)}$} )$ \\ 
$n_2-\bar n_3$ & 0.0045GeV$^3$ & $\frac{1}{3}$
                 & 11.5MeV $\mbox{\boldmath $\sigma^{(2)}$}\cdot \mbox{\boldmath $\sigma^{(3)}$}$ \\ 
$c_1$-$\bar n_3$, $n_2$-$\bar c_4$ & 0.0067GeV$^3$
 & $\frac{1}{3}$ & 4.2MeV$( \mbox{\boldmath $\sigma^{(1)}$}\cdot \mbox{\boldmath $\sigma^{(3)}$}
                   + \mbox{\boldmath $\sigma^{(2)}$}\cdot \mbox{\boldmath $\sigma^{(4)}$} )$ \\ 
$c_1-\bar c_4$ & 0.0123GeV$^3$ & $\frac{1}{3}$
                 & 2.0MeV $\mbox{\boldmath $\sigma^{(1)}$}\cdot \mbox{\boldmath $\sigma^{(4)}$}$ \\ 
\hline
\end{tabular}
\caption{Strength of spin-spin interaction between the constituents.}
\label{tab2}
\end{table}
By using Table \ref{tab2}, we can determine the mass-shifts 
$\Delta M_{SS}$ in ground-states of $cn$-$\bar n\bar c$ system. 
The $J^{PC}$ of $X(3872)$ is plausibly $1^{++}$\cite{Yabsley},
which is obtained by the combinations $\Gamma_{cn}^{0^-}$-$\Gamma_{\bar n\bar c}^{1^-}$
and $\Gamma_{cn}^{1^-}$-$\Gamma_{\bar n\bar c}^{0^-}$.
The $X(3872)$ is used as input to fix overall mass-shift, and
the masses of all the other states are predicted in Table \ref{tab3}. 
\begin{table}
\begin{tabular}{l|cc|c|ccc}
type   & $J^{PC}$ & $\Delta M_{SS}$ & Mass & \multicolumn{3}{c}{decay channels}\\ 
\hline
$\Gamma_{cn}^{0^-}$-$\Gamma_{\bar n\bar c}^{0^-}$ & $0^{++}$ & -65 & 3824 
            & $(J/\psi \rho )$, & $\eta_c \pi$, & $D\bar D$\\
$\frac{1}{\sqrt 2}( \Gamma_{cn}^{0^-}$-$\Gamma_{\bar n\bar c}^{1^-}
                  + \Gamma_{cn}^{1^-}$-$\Gamma_{\bar n\bar c}^{0^-} )$
                                                  & $1^{+-}$ & -27 & 3862
            & $J/\psi\pi$, & $\eta_c\rho$ & \\
$\frac{1}{\sqrt 2}( \Gamma_{cn}^{0^-}$-$\Gamma_{\bar n\bar c}^{1^-}
                  - \Gamma_{cn}^{1^-}$-$\Gamma_{\bar n\bar c}^{0^-} )$
                                                  & $1^{++}$ & -17 & \underline{3872}
            & $J/\psi\rho$ &  & $D^0\bar D^{*0}$ \\
$\Gamma_{cn}^{1^-}$-$\Gamma_{\bar n\bar c}^{1^-}$ & $0^{++}$ & -22 & 3867 
            & $(J/\psi\rho )$, & $\eta_c\pi$, & $D\bar D$\\
$\Gamma_{cn}^{1^-}$-$\Gamma_{\bar n\bar c}^{1^-}$ & $1^{+-}$ &   0.& 3889
            & $J/\psi\pi$, & $\eta_c\rho$, & $D\bar D^*(D$-wave)\\
$\Gamma_{cn}^{1^-}$-$\Gamma_{\bar n\bar c}^{1^-}$ & $2^{++}$ & +43 & 3932
            & $J/\psi\rho$, & & $D\bar D^{(*)}(D$-wave)\\
\hline
\end{tabular}
\caption{Masses of ground-state $cn$-$\bar n\bar c$ system (in MeV). 
The mass of $X(3872)$ is used as input. The $\Delta M_{SS}$ are mass-shifts 
from spin-spin interaction.  
Possible decay channels of $I$=1 states are also shown. For $I$=0 states, $\rho (\pi )$
is replaced with $\omega (\eta )$.
The brackets of $(J/\psi \rho)$ for two $0^{++}$ states, 
of which masses are smaller than $M_\psi +m_\rho$, mean that 
this channel is possible as $J/\psi\pi\pi$ 
through the wide width of $\rho$-meson.
The $\Gamma_{\bar n \bar c}$ are related with $\Gamma_{cn}$ by the charge conjugation transformation, 
$\Gamma_{cn} \rightarrow C^\dagger\ {}^t \Gamma_{cn} C$.
The $C^\dagger\ {}^t\Gamma_{cn}^{0^+,1^+} C= \mp \Gamma_{\bar n \bar c}^{0^+,1^+}$, while
the $C^\dagger\ {}^t\Gamma_{cn}^{0^-,1^-} C= \pm \Gamma_{\bar n \bar c}^{0^-,1^-}$. 
These relations are used to define the charge conjugation parity of total $cn$-$\bar n\bar c$ system.
}
\label{tab3}
\end{table}
The mass of $2^{++}$ state, 
made from $\Gamma_{cn}^{1^-}$-$\Gamma_{\bar n \bar c}^{1^-}$, 
is predicted with 3932MeV, which is very close to the experimental value of $Y(3940)$, 
$M=3943\pm 11 \pm 13$MeV\cite{Y}.
This $2^{++}$ is a candidate of $Y(3940)$. At the same time, 
as well as $X(3872)$ and $Y(3940)$, we predict the existence of the other 4 states
with $J^{PC}=0^{++}$ and $1^{+-}$ in the same mass region. 
However, these extra 4 states are expected to be not observed as narrow-width resonances
but as backgrounds, as will be discussed in the next section.

\section{Strong Decays}

({\it Overlapping of spinor WF})\ \ \ 
The decay mechanism of the relevant quark-rearrangement processes is unknown, however,
it may be reasonable to assume that the decay matrix elements are proportional to the
overlappings of the initial and final WFs. 
The overlapping of spinor WFs, denoted as $A_s$, are given for the decays to 
$c\bar n+n\bar c$ and $c\bar c+n\bar n$, respectively, by 
\begin{eqnarray}
A_s (c\bar n + n\bar c) &=& \langle \Gamma_{\bar n\bar c}^\chi 
     \overline{H}_{\bar D}(v_{\bar D})(-iv_{\bar D}\cdot\gamma)\
    {}^t\Gamma_{cn}^\chi\  {}^t[iv_D\cdot\gamma \overline{H}_D(v_D) ] \rangle\nonumber\\
  &=& \langle \Gamma_{\bar n\bar c}^\chi \ 
    B(v_{\bar D}) \overline{H}_{\bar D}  \bar B(v_{\bar D})\
    {}^t\Gamma_{cn}^\chi\  {}^t[ B(v_D) \overline{H}_D \bar B(v_D) ] \rangle
 \nonumber\\
A_s(c\bar c + n\bar n) &=& \langle \Gamma_{\bar n\bar c}^\chi 
    \overline{H}_{\psi}(v_\psi)\Gamma_{cn}^\chi\ 
    {}^t[iv_\rho\cdot\gamma \overline{H}_\rho (v_\rho) (-iv_\rho\cdot\gamma) ] \rangle \nonumber\\
 &=& \langle \Gamma_{\bar n\bar c}^\chi \ 
    B(v_\psi)\overline{H}_{\psi}\bar B(v_\psi)\ \Gamma_{cn}^\chi\ 
    {}^t[B(v_\rho)\overline{H}_\rho  \bar B(v_\rho) ] \rangle ,
\label{eq10}
\end{eqnarray}
where $D\bar D$ and $J/\psi \rho$ are taken as examples of 
$c\bar n+n\bar c$ and $c\bar c + n\bar n$.
The $\overline{H}(v)$ is the spinor WF of the final meson with velocity $v_\mu (\equiv P_\mu /M)$, 
and $\overline{H}$ is the one at the rest frame with velocity $v_{0\mu}=(0,0,0;i)$. 
They are related by Lorentz booster 
$B(v),\bar B(v)=ch\theta \pm \rho_1\mbox{\boldmath $n$}\cdot\mbox{\boldmath $\sigma$}sh\theta$ as
$\overline{H}(v)=B(v)\overline{H}\bar B(v)$, where the $ch\theta ,sh\theta =\sqrt{\frac{E\pm M}{2M}}$.
In Eq.~(\ref{eq10}), the change of $\rho_3$ on each quark-line in the decay
occurs only through $\rho_1$ term
in $B(v)$ and $\bar B(v)$, which is proportional to $sh\theta$. It is related with the momentum 
${\bf P}$ of the final meson as $ch\theta sh\theta =\frac{\bf P}{2M}$, which is  
much smaller than unity for the relevant decays because of their small phase space. 
Thus, the $\rho_3$ quantum number along every the spectator-quark line is approximately 
conserved\cite{baryon}.

In the chiral-chiral states, the $n$ in $cn$ di-quark and 
the $\bar n$ in $\bar n\bar c$ di-antiquark are in negative $\rho_3$ state. Thus,  
their decay amplitudes to the Pauli mesons, of which $\rho_3$ of constituents are all positive, 
are doubly suppressed with the factor $(\frac{\bf P}{2M})^2$.
In $\tilde U(12)$-scheme,
the WF of vector mesons\cite{CLC}, $J/\psi$, $\rho$, $\omega$, $D^*$ and $\bar D^*$, are 
commonly 
given by $\overline{H}=(1+iv_0\cdot\gamma)\frac{i\epsilon\cdot\gamma}{2\sqrt 2}$, where
the $\rho_3$ of constituents are all positive, $(\rho_3^q,\rho_3^{\bar q})=(+,+)$. 
On the other hand, the WF of pseudoscalar mesons, $\pi$ and $\eta$, which are Nambu-Goldstone bosons,
are $\overline{H}=\frac{i\gamma_5}{2}$, which has doubly-negative $\rho_3$ component as 
$(\rho_3^q,\rho_3^{\bar q})=\frac{1}{\sqrt 2}(\ (+,+)+(-,-)\ )$(, while the WF of $D$ and $\eta_c$ is 
$\overline{H}=(1+iv_0\cdot\gamma)\frac{\gamma_5}{2\sqrt 2}$, which has  
$(\rho_3^c,\rho_3^{\bar c})=(+,+)$). 
Thus, in the relevant decays of the chiral-chiral states,
the transition amplitudes to $J/\psi \rho (\omega )$, $\eta_c \rho (\omega)$ and $D\bar D^{(*)}$
are strongly suppressed from $\rho_3$-line rule, while the ones to $J/\psi \pi (\eta)$
and $\eta_c \pi (\eta)$ are not suppressed. 
The decay widths of the latter channels are expected to be a few hundred MeV or a GeV.

Thus, when the initial states have no $\pi (\eta)$ decay modes, they have narrow
widths, and are expected to be observed as resonant particles. 
The possible decay channels of the ground six states are shown in Table \ref{tab3}.
{\it Among them, only two states with $J^{PC}=1^{++}$ and $2^{++}$ have no $\pi (\eta)$ decay modes
and are considered to be detected as resonances.} 
Concerning these two states the decays to $\eta_c \pi (\eta)$
are also forbidden from the conservation of heavy quark spin\cite{Voloshin}.
This theoretical expectation is consistent with the present experimetnal situation:
Only two states, $X(3872)$ and $Y(3940)$, are observed as the exotic candidates in the
relevant mass region.

({\it Decay widths})\ \ \ 
We can also calculate the overlappings of space WFs, denoted as $\tilde F_x$. 
%
%
They are given by
\begin{eqnarray}
\tilde F_x({c\bar n + n\bar c}) &=& \int d^3{\bf p}_1 d^3{\bf p}_2 f_D({\bf q}_D) f_{\bar D}({\bf q}_{\bar D})
                     \cdot  f_s({\bf p}_s) f_R({\bf p}_R) f_U({\bf p}_U) ,\nonumber\\ 
\tilde F_x({c\bar c + n\bar n}) &=& \int d^3{\bf p}_1 d^3{\bf p}_2 
                       f_\psi({\bf q}_\psi) f_{\rho}({\bf q}_{\rho})
                     \cdot  f_s({\bf p}_s) f_R({\bf p}_R) f_U({\bf p}_U)\ \ ,
\label{eq8}
\end{eqnarray}
where
the ${\bf p}_s$ etc. are internal momenta,  
the $f_s({\bf p}_s)$'s are WFs in momentum representation,
$f_s({\bf p}_s)=(\frac{1}{\pi\beta_s})^{\frac{3}{4}}e^{-\frac{1}{2\beta_s}{\bf p}_s^2}$.
%
Their numerical values are $\tilde F_x(c\bar n + n\bar c)=0.96$GeV$^{-\frac{3}{2}}$
and $\tilde F_x(c\bar c + n\bar n)=1.05$GeV$^{-\frac{3}{2}}$, respectively.
We neglect the difference between them, and use a common value $\tilde F_x\simeq 1$GeV$^{-\frac{3}{2}}$.

The width $\Gamma$ of the decay $X$(or $Y$)$\rightarrow A+B$ is given by 
\begin{eqnarray}
\Gamma &=&  a |{\bf P}|  ( E_A E_B M \tilde F_x^2 )  | A_s |^2
\label{eq9}
\end{eqnarray}
with one dimension-less parameter $a$, where $M$ is the mass of initial 4 quark state, and 
$E_A(E_B)$ is the energy of the final meson $A(B)$. 
The $|{\bf P}|$ is the momentum of final meson.
The explicit forms of $A_s$ and $|A_s|^2$, and the corresponding decay widths for 
the relevant decay modes of $X(3872)$ and $Y(3940)$ are given in 
Table \ref{tab4}. The parameter $a$ is fixed with $a=18.8$, which is determined by the 
the total width of $Y(3940)$, $\Gamma_Y=87$MeV, as input.  
\begin{table}
\begin{tabular}{l|ll|l}
channel & $A_s$ & $\overline{|A_s|^2}$ & $\Gamma$ in MeV\\
\hline
$Y(3940)\rightarrow J/\psi\omega$ & 
   $- \tilde\epsilon^\psi_i \ \epsilon_{ij} \  \tilde\epsilon^\omega_j$ sh$^2\theta_\omega$
    & $(\frac{{\bf p}_\omega}{2m_\omega})^4$ & 60.3\ cos$^2\theta_Y$\\
$Y(3940)\rightarrow D\bar D$ &
   $- n^D_i \ \epsilon_{ij} \  n^D_j$ \ $\frac{{\bf P}_D^2}{4M_D^2}$
    & $(\frac{{\bf P}_D}{2 M_D})^4\ \frac{2}{15}$ & 20.1\\
$Y(3940)\rightarrow D\bar D^*$ &
   $i (\mbox{\boldmath $n$}^{\bar D^*}$$\times$$\tilde{\mbox{\boldmath $\epsilon$}}^{\bar D^*})_i
     \epsilon_{ij} n^{\bar D^*}_j$\ $\frac{{\bf P}_D^2}{4M_DM_{\bar D^*}}$
    & $(\frac{{\bf P}_D^2}{4 M_DM_{D^*}})^2\ \frac{2}{5}$ & 6.6\\
\hline
$X(3872)\rightarrow J/\psi \rho$ &
   $-\frac{i}{\sqrt 2}
   (\mbox{\boldmath $\epsilon$}$$\times$$\tilde{\mbox{\boldmath $\epsilon$}}^{\psi})
    \cdot \tilde{\mbox{\boldmath $\epsilon$}}^{\rho}$\ sh$^2\theta_\rho$  
    & $(\frac{{\bf p}_\rho}{2m_\rho})^4$ & (1.2 sin$^2\theta_X$) \\
$X(3872)\rightarrow J/\psi \omega$ & & $(\frac{{\bf p}_\omega}{2m_\omega})^4$ & (0.2 cos$^2\theta_X$) \\
$X(3872)\rightarrow D^0\bar D^{*0}$ &
   $\epsilon_i n^D_i n^D_j \tilde \epsilon^{\bar D^*}_j$\ 
   $\frac{\sqrt 2\ {\bf P}_D^2}{4M_DM_{\bar D^*}}$
    & $(\frac{{\bf P}_D^2}{4 M_DM_{D^*}})^2\ \frac{2}{3}$ & $\sim 0$(1.5$\times$$10^{-4}$)\\
\hline
\end{tabular}
\caption{Decay widths $\Gamma$ of $Y(3940)$ and $X(3872)$. 
The parameter $a$(see, Eq.~(\ref{eq9})) is fixed with $a=18.8$ through $\Gamma_Y^{\rm tot}$=87MeV
as input.
The overlappings of spinor WF, $A_s$, and their spin-averaged squares
$\overline{|A_s|^2}$ are also given.
The $\epsilon_{ij}(\mbox{\boldmath $\epsilon$})$ is the $J$=2(1) polarization-tensor of $Y(X)$.
The $\mbox{\boldmath $n$}^{D,\bar D^*}$ are unit vectors to the directions of $D,\bar D^*$ momenta.
For the $\Gamma$ of $Y\rightarrow D\bar D^*$, $\bar D D^*$ is also included. 
The cos$\theta_{X,Y}$(sin$\theta_{X,Y}$) are $I$=0(1) component of the flavor WFs of $X,Y$.
The $\Gamma (X\rightarrow J/\psi\omega )$, of which decay mode is above the threshold, 
is given only by multiplying $ \langle {\bf p}_\omega \rangle / \langle {\bf p}_\rho \rangle$ 
with $\Gamma (X\rightarrow J/\psi\rho)$.
}
\label{tab4}
\end{table}
The partial decay rates of $Y(3940)$
are fairly consistent with the present experiments. 
By taking the $I$=0 component cos$\theta_Y \simeq 1$, the main decay mode of $Y(3940)$ is considered
to be $J/\psi\omega$.
Experimentally the $Y(3940)$ is observed in $J/\psi\omega$, not in $D\bar D$ and $D\bar D^*$.
The $\Gamma$ of $X$ are largely dependent upon the treatment of 
the effective momenta of final mesons.
In the case of $X\rightarrow J/\psi\rho$, it is replaced with the effective momentum
$\langle{\bf p}_\rho\rangle$, defined by
$\langle{\bf p}_\rho\rangle=\int_{4m_\pi^2}^{(M_X-M_{\psi})^2} \frac{m_\rho\Gamma_\rho}{\pi}
 \frac{\sqrt{(M_X^2-M_\psi^2-s)^2-4sM_\psi^2}/(2M_X)}{|m_\rho^2-s-im_\rho\Gamma_\rho|^2}=132$MeV.
The results are shown with parentheses.
However, if we take sin$^2\theta_X$$\simeq$1$/6$,
their decay properties are also consistent with the present experiments,
$\Gamma_X^{\rm tot}<2.3$MeV\cite{X} and 
$\frac{\Gamma (X\rightarrow J/\psi\ 3\pi)}{\Gamma (X\rightarrow J/\psi\ 2\pi )}
(\simeq   \frac{\Gamma (X\rightarrow J/\psi\ \omega)}{\Gamma (X\rightarrow J/\psi\ \rho )} ) $
$=$1.0$\pm$0.4$\pm$0.3.\cite{BelleGam}

\section{Concluding Remarks}

The $X(3872)$ and $Y(3940)$ have the properties of 
the four quark $cn$-$\bar n\bar c$ states with $J^{PC}=1^{++}$ and $2^{++}$, respectively,
which come from $0^-$-$1^-$ and $1^-$-$1^-$ combinations of 
$cn$ chiral di-quark and $\bar n\bar c$ chiral di-antiquark in $\tilde U(12)$-scheme.
Their mass difference is consistently explained by the spin-spin interaction between constituents
in JSQM.
Their narrow widths are explained from a phenomenological selection rule, 
coming from an orthogonality of spinor WFs, called $\rho_3$-line rule.
Only these two states are expected to be observed as resonant particles in ground states, and the 
other states have very large widths and are not observed as resonances but as non-resonant backgrounds.
\begin{table}
\begin{tabular}{c|c|l}
state & Mass(MeV) & $\Gamma$(channel)\\
\hline
$X_{cs-\bar s\bar c}^{1^{++}}$ & $\sim$4080 & $\Gamma^{\rm tot}<1$MeV  \\
\hline
$Y_{cs-\bar s\bar c}^{2^{++}}$ & $\sim$4150 & $\Gamma (Y^{2^{++}}$$\rightarrow$$D_s \bar D_s)$=23MeV \\
 & & $\Gamma (Y^{2^{++}}$$\rightarrow$$D_s \bar D_s^*)$=8MeV \\
 & & $\Gamma (Y^{2^{++}}$$\rightarrow$$J/\psi \phi )$=11MeV \\
\hline
\end{tabular}
\caption{The predictions for the ground $cs$-$\bar s\bar c$ states. The $J^{PC}=1^{++}$ and $2^{++}$
states, denoted, respectively, as $X_{cs-\bar s\bar c}^{1^{++}}$ and $Y_{cs-\bar s\bar c}^{2^{++}}$, 
are expected to be observed. 
Their masses and the partial decay widths of $Y^{2^{++}}_{cs-\bar s\bar c}$ are given. 
Its total width is simply estimated by the sum as 
$\Gamma^{\rm tot}_{Y^{2^{++}}_{cs-\bar s\bar c}}$=23+8+11=42MeV.
 }
\label{tab5}
\end{table}

In order to check this interpretation,
we present the predictions for the properties of ground $cs$-$\bar s\bar c$ states in Table \ref{tab5}. 
Their masses are estimated simply by adding $2(M_{D_s^*}-M_{D^{*0}})\simeq 210$MeV
to the corresponding $cn$-$\bar n\bar c$ states. 
Their decay widths are estimated by using Eq.~(\ref{eq9}) 
with the common parameter $a$.
The mass of $X_{cs-\bar s\bar c}^{1^{++}}$ is expected to be quite close
to the $D_s\bar D_s^*$ threshold and its decay is strongly suppressed.
The $J/\psi \phi$ is un-open channel. Thus, the main decay mode is considered to be 
electromagnetic ones of order keV.
The main decay mode of $Y_{cs\bar s\bar c}^{2^{++}}$ is $D_s\bar D_s$. 
In order to check the four quark nature of this state, 
it is necessary to observe $J/\psi\phi$ decay.\\
{\it The authors would like to express their sincere gratitudes to the members of $\sigma$-group
for useful comments and encouragements. M. I. is very grateful to professor M. Oka for
valuable comments and discussions.}

\end{document}